\documentclass[showpacs,preprintnumbers,amsmath,prd,amssymb,superscriptaddress]{revtex4}
\usepackage{amssymb}
\usepackage{CJK}
\usepackage{amsmath,amssymb,graphicx,bm}

\begin{document}

\title{Conformal transformation in $f(T)$ theories}

\author{Rong-Jia Yang}
\email{yangrj08@gmail.com}
 \affiliation{College of Physical Science and Technology,
Hebei University, Baoding 071002, China}

\begin{abstract}
It is well-known that $f(R)$ theories are dynamically equivalent to a particular class of scalar-tensor theories. In analogy to the $f(R)$ extension of the Einstein-Hilbert action of general relativity, $f(T)$ theories are generalizations of the action of teleparallel gravity. The field equations are always second order, remarkably simpler than $f(R)$ theories. It is interesting to investigate whether $f(T)$ theories have the similar conformal features possessed in $f(R)$ theories. It is shown, however, that $f(T)$ theories are not dynamically equivalent to teleparallel action plus a scalar field via conformal transformation, there appears an additional scalar-torsion coupling term. We discuss briefly what constraint of this coupling term may be put on $f(T)$ theories from observations of the solar system.
\end{abstract}

\makeatletter
\def\@pacs@name{PACS: 04.50.Kd, 04.50.-h, 95.36.+x, 11.25.Hf}
\makeatother

\maketitle

\section{Introduction}
The dynamical equivalence between $f(R)$ theories (see e. g. \cite{Capozziello06, Nojiri03,Nojiri07,Nojiri10}) and a particular class of scalar-tensor (ST) theories, is very well-known and studied, both in the case of metric formalism \cite{Futamase}, as well as in the Palatini formalism \cite{Fla}.
More precisely, one can demonstrate that a ST theory action is conformally equivalent to the Einstein-Hilbert action plus as many scalar fields as there are in the former action. On the other hand, when a conformal transformation of an $f(R)$ action from the starting frame (referred to as ``Jordan frame") to the final one (called ``Einstein frame") is performed, one finds that it is equivalent to Einstein-Hilbert action plus a scalar field. See, for example, for an action of $f(R)$ in terms of metric,
\begin{eqnarray}\label{R}
I=\frac{1}{2k^2}\int d^4x \sqrt{-g} f(R)+I_{\rm m}(g_{\mu \nu}),
\end{eqnarray}
where $k^2=8\pi G$, the field equation is given by taking variations with respect to the metric. It is easy to see, for a given function $f(\phi)$, provided that $f''(\phi)\neq 0$, the following action leads to the same field equation,
\begin{eqnarray}\label{R1}
I= \frac{1}{2k^2} \int d^4x \sqrt{-g} [f(\phi)+(R-\phi)f'(\phi)]+I_{\rm m}(g_{\mu \nu}),
\end{eqnarray}
where $\phi$ is an auxiliary field and $f'(\phi)=df/d\phi$. The action (\ref{R1}) can be transformed into a general ST action,
\begin{eqnarray}\label{R2}
I_{\rm JF}= \frac{1}{2k^2} \int d^4x \sqrt{-g} [F(\varphi)R-\omega(\varphi)g^{\mu\nu}\nabla_\mu \varphi \nabla_\nu \varphi-2V(\varphi)]+I_{\rm m}(g_{\mu \nu}),
\end{eqnarray}
by setting in the latter the identifications: $F(\varphi)=f'(\phi)$, $\omega(\varphi)=0$, and $2V(\varphi)=\phi f'(\phi)-f(\phi)$. That is to say, action (\ref{R})
in the metric formalism can be identified with the case $\omega=0$ of the Brans-Dicke (BD) theory. Make the conformal transformation: $\hat{g}_{\mu\nu}=\Omega^2g_{\mu\nu}$,
the action (\ref{R2}) takes the form of the Einstein-Hilbert action plus a scalar field,
\begin{eqnarray}\label{R3}
I_{\rm EF}= \frac{2}{k^2} \int d^4x \sqrt{-\hat{g}} \left[\frac{\hat{R}}{4}-\frac{1}{2}\hat{g}^{\mu\nu}\hat{\nabla}_\mu \psi \hat{\nabla}_\nu \psi-U(\psi) \right]+I_{\rm m}[ F^{-1}(\varphi)\hat{g}_{\mu\nu}],
\end{eqnarray}
with $F(\varphi)=\Omega^2(x)$, $2U(\psi)\equiv V(\varphi)F^{-2}(\varphi)$, and $(d\psi/d\varphi)^2\equiv \frac{3}{4} (d\ln F(\varphi)/d\varphi)^2+\omega(\varphi)/2F(\varphi)$. The matter term is now non-minimally coupled to the scalar $\psi$ through the conformal factor $F^{-1}(\varphi)$. The dynamical equivalence between BD theory and $f(R)$ theories (see e. g. \cite{Nojiri,Capone} and references therein) suggests to use the results known for the former to directly obtain, after suitable manipulations, those corresponding to the latter. This could be very fruitful especially for those results directly related to observations or experiments: for instance, post-Newtonian parameters (see \cite{Will} for a recent review) can be used to constrain $f(R)$ theories.

Recently, models based on modified teleparallel gravity were proposed as an alternative to $f(R)$ theories \cite{Bengochea}, namely $f(T)$ theories, in which the torsion will be responsible for the late accelerated expansion, and the field equations will always be 2nd order equations, remarkably simpler than $f(R)$ theories. This feature has led to a rapidly increasing interest in the literature (see, for some recent work \cite{Linder, Myrzakulov, Yerzhanov,yang, Li, Dent,Bamba,Bengochea1,Zheng,wu1}). \textbf{Since some new types of $f(T)$ theories have been constantly proposed \cite{Linder, Myrzakulov, Yerzhanov, yang}, in analogy to $f(R)$ theories, it is valuable to investigate whether $f(T)$ theories have the similar conformal feature possessed in $f(R)$ theories. In other words, whether $f(T)$ theories and ST theories are dynamically equivalent. If it is true, then one can use the results obtained from ST theories to directly put the observational constraints on $f(T)$ theories, for instance, the post-Newtonian parameters know for ST theories can be used to constrain $f(T)$ theories. If it is not true, what new effects can be presented when looking at the observational consequences of $f(T)$ theories. This is the main goals of this work.}

The paper is organized as follows, in the following section, we review $f(T)$ theories. In Sec. III, we consider the conformal transformations in $f(T)$ theories. Finally, we shall close with a few concluding remarks in Sec. IV.

\section{$f(T)$ gravities}
Rather than use the curvature defined via the Levi-Civita connection, one could explore the torsion via the Weitzenb\"{o}ck connection that has no curvature to establish theories of gravity
\begin{align}
\label{torsion2}
T^\lambda_{\:\:\:\mu\nu}\equiv e^\lambda_i(\partial_\mu e^i_\nu-\partial_\nu e^i_\mu),
\end{align}
where $e_i^\mu$ ($\mu=0, 1, 2, 3$) are the components of the vierbein field ${\mathbf{e}_i(x^\mu)}$ ($i=0, 1, 2, 3$) in a coordinate basis, i.e. $\mathbf{e}_i=e^\mu_i\partial_\mu$.
The vierbein is an orthonormal basis
for the tangent space at each point $x^\mu$ of the manifold: $\mathbf{e}%
_i\cdot\mathbf{e}_j=\eta_{i\, j}$, where $\eta_{i\, j}=$diag $(1,-1,-1,-1)$. Notice that Latin indexes refer to the tangent space, while Greek indexes
label coordinates on the manifold. The metric tensor is obtained from the
dual vierbein as $g_{\mu\nu}(x)=\eta_{i\, j}\, e^i_\mu (x)\, e^j_\nu (x)$. This approach (named ``teleparallelism") by using the vierbein as dynamical object was taken by Einstein \cite{Einstein}.

The teleparallel Lagrangian is \cite{Hayashi, Maluf, Arcos},
\begin{equation}  \label{lagTele}
T\equiv S_\rho^{\:\:\:\mu\nu}\:T^\rho_{\:\:\:\mu\nu}=\frac{1}{4}T_{\rho}^{\:\:\:\:\mu\nu}T^\rho_{\:\:\mu\nu}-\frac{1}{2}T^{\mu\nu}_{\:\:\:\:\rho}T^\rho_{\:\:\mu\nu}
+T^{\theta\mu}_{\:\:\:\:\theta}T^\rho_{\:\:\rho\mu},
\end{equation}
where:
\begin{equation}  \label{S}
S_\rho^{\:\:\:\mu\nu}=\frac{1}{2}\Big(K^{\mu\nu}_{\:\:\:\:\rho}+\delta^\mu_%
\rho \:T^{\theta\nu}_{\:\:\:\:\theta}-\delta^\nu_\rho\:
T^{\theta\mu}_{\:\:\:\:\theta}\Big),
\end{equation}
and the contorsion tensor, $K^{\mu\nu}_{\:\:\:\:\rho}$, is
\begin{equation}  \label{K}
K^{\mu\nu}_{\:\:\:\:\rho}=-\frac{1}{2}\Big(T^{\mu\nu}_{\:\:\:\:\rho}
-T^{\nu\mu}_{\:\:\:\:\rho}-T_{\rho}^{\:\:\:\:\mu\nu}\Big),
\end{equation}
which equals the difference between Weitzenb\"{o}ck and Levi-Civita connections.

Following Ref. \cite{Bengochea} we promote the teleparallel Lagrangian density as a function of $T$, in analogy to $f(R)$ theories.
Thus the action reads
\begin{eqnarray}\label{f}
I= \frac{1}{2k^2} \int d^4x\:e\:f(T)+I_{\rm m}(e^i_\mu),
\end{eqnarray}
where $e=det(e^i_\mu)=\sqrt{-g}$. If matter couples to the metric in the standard form then the variation of
the action with respect to the vierbein leads to the equations
\begin{eqnarray}
[e^{-1}\partial_\mu(e\:S_i^{\:\:\:\mu\nu})-e_i^{\:\lambda}
\:T^\rho_{\:\:\:\mu\lambda}\:S_\rho^{\:\:\:\nu\mu}]f_T+ S_i^{\:\:\:\mu\nu}\partial_\mu T f_{TT}+\frac{1}{4}%
\:e_i^\nu \:f(T)=\frac{1}{2}k^2\:e_i^{\:\:\:\rho}\:\Theta_\rho^{\:\:\:\nu},  \label{ecsmovim}
\end{eqnarray}
where $f_T\equiv df/dT$, $f_{TT} \equiv d^2f/dT^2$, $S_i^{\:\:\mu\nu}\equiv e_i^{\:\:\rho}S_\rho^{\:\:\mu\nu}$, and
$\Theta_{\mu\nu}$
is the matter energy-momentum tensor. The fact that equations (\ref{ecsmovim}) are 2nd order makes them simpler than the dynamical equations resulting in
$f(R)$ theories. Assuming a flat homogeneous and isotropic FRW universe, one has $T\equiv S^{\rho\mu\nu}T_{\rho\mu\nu}=-6\:H^2$, where $H$ is the Hubble parameter $H=\dot{a}/a$. The modified Friedmann equations read as
\begin{eqnarray}\label{f1}
	12H^2 f_{T}+f &=& 2k^2 \rho,\\
\label{f2}
	48 H^2 \dot{H}f_{TT}-(12H^2+4\dot{H})f_{T}-f &=& 2k^2p,
\end{eqnarray}
where $\rho$ and $p$ are the total density and pressure respectively. The torsion contributions to the energy density and pressure are, respectively \cite{yang,Myrzakulov,Yerzhanov},
\begin{eqnarray}
\label{rhoT}
\rho_T &=& \frac{1}{2k^2}(-12H^2f_T-f+6H^2), \\
\label{pT}
p_T &=& -\frac{1}{2k^2}[48\dot{H}H^2f_{TT}-4\dot{H}f_{T}+4\dot{H}]-\rho_T,
\end{eqnarray}

\section{Conformal transformations in $f(T)$ theories}
The action (\ref{f}) can be generalized as:
\begin{eqnarray}\label{f1}
I= \frac{1}{2k^2} \int d^4x e [f(\phi)+(T-\phi)f'(\phi)]+I_{\rm m}(e^i_\mu),
\end{eqnarray}
where $f'(\phi)\equiv df/d\phi$. One can easily verify that the field equation ($(T-\phi)f''(\phi)=0$) for $\phi$ gives $\phi=T$ if $f''(\phi)\neq 0$, which reproduces the action (\ref{f}). Introducing a scalar field $\varphi$ such that $F(\varphi)=f'(\phi)$, the action (\ref{f1}) can be written as:
\begin{eqnarray}\label{f2}
I_{\rm JF}= \frac{1}{2k^2} \int d^4x e [F(\varphi)T-\omega(\varphi)g^{\mu\nu}\nabla_\mu \varphi \nabla_\nu \varphi-2V(\varphi)]+I_{\rm m}(e^i_\mu),
\end{eqnarray}
where $2V(\varphi)=\phi f'(\phi)-f(\phi)$. The action (\ref{f}) corresponds to a BD theory with $\omega=0$ in Jordan frame. It was shown above that an $f(R)$ action is equivalent to Einstein-Hilbert plus a scalar field action via conformal transformation. It is interesting to investigate whether $f(T)$ theories have the similar feature of $f(R)$ theories under a conformal transformation.

Considering a conformal transformation of the metric, defined as:
\begin{eqnarray}\label{ct}
\hat{g}_{\mu\nu}=\Omega^2(x)g_{\mu\nu},
\end{eqnarray}
where $\Omega(x)$ is a smooth, non-vanishing function of the space-time point, lying in the range $0<\Omega< \infty$. \textbf{One must be careful, however, that such transformations may break down at singular points, like the case in  $F(R)$ gravity \cite{Briscese,Bamba08}}. We will use a caret to indicate
quantities in the transformed frame. From Eq. (\ref{ct}), we immediately see that,

\begin{eqnarray}\label{ct1}
\hat{g}^{\mu\nu}=\Omega^{-2}(x)g^{\mu\nu}, ~~~~\hat{e}^{a}_\mu=\Omega(x)e^{a}_\mu,~~~~~\hat{e}^{\mu}_a=\Omega^{-1}(x)e^{\mu}_a,~~~~ \hat{e}=\Omega^4e
\end{eqnarray}

Under the transformation of Eq. (\ref{ct}), one can find the torsion transforms as:
\begin{eqnarray}
\hat{T}^\rho_{\:\:\:\mu\nu}=T^\rho_{\:\:\:\mu\nu}+\Omega^{-1}[\delta^\rho_\nu \partial_\mu \Omega-\delta^\rho_\mu \partial_\nu \Omega],
\end{eqnarray}
$S_\rho^{\:\:\:\mu\nu}$ transforms as:
\begin{equation}
\hat{S}_\rho^{\:\:\:\mu\nu}=\frac{1}{2}\Big(\hat{K}^{\mu\nu}_{\:\:\:\:\rho}+\hat{\delta}^\mu_
\rho \:\hat{T}^{\theta\nu}_{\:\:\:\:\theta}-\hat{\delta}^\nu_\rho\:
\hat{T}^{\theta\mu}_{\:\:\:\:\theta}\Big)=\Omega^{-2}S_\rho^{\:\:\:\mu\nu}+\Omega^{-3}(\delta^\mu_\rho \partial^\nu \Omega-\delta^\nu_\rho \partial^\mu \Omega),
\end{equation}
and the teleparallel Lagrangian transforms as:
\begin{eqnarray}
\hat{T}=\Omega^{-2}T+4\Omega^{-3}\partial^\mu \Omega T^\rho_{\:\:\:\rho\mu}-6\Omega^{-4} \partial_\mu \Omega\partial^\mu \Omega,
\end{eqnarray}
with the inverse transformation is given by,
\begin{eqnarray}\label{T}
T= \Omega^{2} \hat{T}-4\Omega^{-1}\partial^\mu \Omega \hat{T}^\rho_{\:\:\:\rho\mu}+6\Omega^{-2} \partial_\mu \Omega\partial^\mu \Omega
\end{eqnarray}
One must be careful to specify whether one is taking derivatives with respect to the original vierbein $e^{a}_\mu$ or the transformed vierbein $\hat{e}^{a}_\mu$, because the Weitzenb\"{o}ck connection (and hence covariant derivatives) transform in $\Omega$-dependent ways under the transformation of Eq. (\ref{ct}). Recall that $x^\mu$ is unaffected by the conformal transformation, so that $\partial_\mu=\hat{\partial}_\mu$. For scalar functions, we have $\nabla_\mu \Omega=\partial_\mu \Omega$, and hence $\nabla_\mu \Omega=\hat{\nabla}_\mu \Omega$.

Using Eqs. (\ref{T}), we may rewrite the action (\ref{f2}) as:
\begin{eqnarray}\label{f3}
I_{\rm EF}= \frac{1}{2k^2} \int d^4x \hat{e} [\hat{T}+2F^{-3}\hat{\partial}^\mu F \hat{T}^\rho_{\:\:\:\rho\mu}-\frac{1}{2}\hat{g}^{\mu\nu}\hat{\nabla}_\mu \psi \hat{\nabla}_\nu \psi-U(\psi)]+I_{\rm m}[F(\varphi)^{-1/2}\hat{e}^i_\mu],
\end{eqnarray}
where $F(\varphi)=\Omega^2$, $U(\psi)=2V(\varphi)/F^2(\varphi)$, and $(d\psi/d\varphi)^2=2\omega/F-3[F'(\varphi)]^2/F^4$. Comparing with action (\ref{R3}), there is an additional scalar-torsion coupling term, $2F^{-3}\hat{\partial}^\mu F \hat{T}^\rho_{\:\:\:\rho\mu}$, in action (\ref{f3}), which cannot be removed by conformal transformation. In other words, $f(T)$ theories are not dynamically equivalent to teleparallel action plus a scalar field via conformal transformation, their conformal features differ from that of $f(R)$ theories. By varying such a transformed action with respect to the vierbein, $\hat{e}_\lambda^a$, we obtain the subsequent equation
for the components of the vierbein
\begin{eqnarray}
&&\hat{G}_a^\lambda=-\frac{1}{2}\hat{\partial}_\mu[\hat{e}F^{-3}(\hat{\partial}^\lambda F\hat{e}^\mu_a- \hat{\partial}^\mu F\hat{e}^\lambda_a)]+ \frac{1}{2}\hat{e}\hat{e}^\lambda_aF^{-3}\hat{\partial}^\mu F \hat{T}^\rho_{\:\:\:\rho\mu}- \frac{1}{2}\hat{e}\hat{e}^\rho_aF^{-3}\hat{\partial}^\mu F \hat{T}^\lambda_{\:\:\:\rho\mu}\\ \nonumber
&&+ \frac{1}{4}\hat{e}\hat{e}^\nu_a \hat{\nabla}^\lambda \psi \hat{\nabla}_\nu \psi- \frac{1}{8}\hat{e}\hat{e}^\lambda_a \hat{\nabla}^\mu \psi \hat{\nabla}_\mu \psi- \frac{1}{4}\hat{e}\hat{e}^\lambda_a U(\psi)+ \frac{1}{2}k^2\hat{e}\hat{e}^\sigma_a \hat{\Theta}_\sigma^\lambda
\end{eqnarray}
where $\hat{G}_a^\lambda\equiv \hat{\partial}_\mu(\hat{e} \hat{S}_a^{\:\:\:\mu\lambda})+ \hat{e}\hat{e}_a^{\:\nu} \:\hat{T}^\rho_{\:\:\:\mu\nu}\:\hat{S}_\rho^{\:\:\:\mu\lambda}- \frac{1}{4}\hat{e}\:\hat{e}_a^\lambda \hat{T}$, corresponding to the Einstein tensor in general relativity. Whereas a variation with respect to the scalar field, $\psi$, gives the equation
\begin{eqnarray}
\label{es}
\hat{\square}\psi=-4\pi G \alpha(\psi) \hat{\Theta}-2\hat{T}^\rho_{\:\:\:\rho\mu}\frac{d(F^{-3}\hat{\partial}^\mu F)}{d\psi} +\frac{dU(\psi)}{d\psi},
\end{eqnarray}
where $\alpha(\psi) \equiv d\ln F^{-1/2}/d\psi$ is a function giving the strength of the coupling between the scalar field and the matter/energy source, and $\hat{\Theta}$ is the trace of the stress-energy tensor. With equation (\ref{es}), we can investigate the effect of the coupling term, $2F^{-3}\hat{\partial}^\mu F \hat{T}^\rho_{\:\:\:\rho\mu}$ (or the term, $-2\hat{T}^\rho_{\:\:\:\rho\mu}d(F^{-3}\hat{\partial}^\mu F)/d\psi$, in Eq. (\ref{es})), putted on $f(T)$ theories from observations of the solar system.

According to the procedure described in \cite{Esposito1}, we can determine the post-Newtonian parameters $\gamma$ and $\beta$ for a $f(T)$ theory as following
\begin{eqnarray}
\label{ppn1}
\gamma-1=-\frac{2\alpha^2}{1+\alpha^2}\big|_{\psi_0},
\end{eqnarray}
and
\begin{eqnarray}
\label{ppn2}
\beta-1=\frac{1}{2}\left[\frac{\alpha^2}{(1+\alpha^2)^2} \frac{d\alpha}{d\psi}\right]_{\psi_0},
\end{eqnarray}
where $\psi_0$ is the asymptotic value of the field $\psi$. According to Eqs. (\ref{ppn1}) and (\ref{ppn2}), it seems that the coupling term, $2F^{-3}\hat{\partial}^\mu F \hat{T}^\rho_{\:\:\:\rho\mu}$, have no effect on the post-Newtonian parameters $\gamma$ and $\beta$. We note, however, the post-Newtonian parameters (\ref{ppn1}) and (\ref{ppn2}) are suitable only when the potential of the scalar field $\psi$ is absolutely negligible (that is to say $F\gg 1$). If the potential can not be neglected, we must take the procedure described in \cite{Capone} or \cite{Perivolaropoulos} to obtain the post-Newtonian parameters. However, these two methods can not tell us what constraint of the coupling term, $2F^{-3}\hat{\partial}^\mu F \hat{T}^\rho_{\:\:\:\rho\mu}$, may be putted on  $f(T)$ theories by using observations of gravity within the solar system, because it is no need to consider a conformal transformation of the metric to obtain the post-Newtonian parameters, that is to say, the coupling term, $2F^{-3}\hat{\partial}^\mu F \hat{T}^\rho_{\:\:\:\rho\mu}$, does not appear in these two methods.

\section{Summary and conclusions}
Recently $f(T)$ theories, based on modifications of the teleparallel gravity where torsion is the geometric object describing gravity instead of curvature, are proposed to explain the cosmic speed-up. The field equations are always second order, remarkably simpler than $f(R)$ theories. In this work, it was shown that $f(T)$ theories are not dynamically equivalent to teleparallel action plus a scalar field under conformal transformation, unlike $f(R)$ theories. An additional scalar-torsion coupling term in the case of $f(T)$ theories cannot be removed by conformal transformation.

For the case in which the potential of the scalar field $\psi$ is absolutely negligible, we obtain the post-Newtonian parameters $\gamma$ and $\beta$. With these post-Newtonian parameters, however, we can not determine what effect of the coupling term, $2F^{-3}\hat{\partial}^\mu F \hat{T}^\rho_{\:\:\:\rho\mu}$, may be putted on the post-Newtonian parameters. If the potential can not be neglected, one can not directly obtain the post-Newtonian parameters from ST theory to constrain $f(T)$ theories, since $f(T)$ theories are not dynamically equivalent to teleparallel action plus a scalar field. It should be find out another way to test $f(T)$ theories with solar-system experiments, such as the methods in \cite{Capone} or \cite{Perivolaropoulos}, but these two methods can not tell us what constraint of the coupling term may be putted on $f(T)$ theories from observations of the solar system, because it is no need to consider a conformal transformation of the metric to obtain the post-Newtonian parameters.

\begin{acknowledgments}
We thank M. Capone for helpful discussions. This study is supported in part by Research Fund for Doctoral
Programs of Hebei University No. 2009-155, and by Open Research
Topics Fund of Key Laboratory of Particle Astrophysics, Institute of
High Energy Physics, Chinese Academy of Sciences, No.
0529410T41-200901.
\end{acknowledgments}

\bibliography{apssamp}

\end{document}